\documentclass[authoryear]{elsarticle}

\usepackage[utf8]{inputenc}
\usepackage{natbib}
\usepackage{url}
\usepackage{hyperref,xcolor}

\usepackage{amsmath}	
\usepackage{amssymb}	

\usepackage{subcaption}
\usepackage[version=4]{mhchem}

\usepackage[margin=1.0in]{geometry}

\def\nat{Nature }

\def\aap{A\&A }
\def\aj{AJ }
\def\icarus{Icarus }
\def\grl{GRL }
\def\planss{Planet. Space Sci. }

\usepackage{setspace}

\begin{document}
\begin{frontmatter}

\author[add1,add7]{Edward Gillen\footnote{These authors contributed equally}}
\author[add2,add1,add3]{Paul B. Rimmer$^1$}
\author[add4]{David C. Catling}
\address[add1]{Cavendish Astrophysics, Cavendish Laboratory, JJ Thomson Ave, Cambridge CB3 0HE. UK}
\address[add2]{Department of Earth Science, Downing St, Cambridge CB2 3EQ. UK}
\address[add3]{Laboratory for Molecular Biology, Francis Crick Ave, Cambridge CB2 0QH. UK}
\address[add4]{Department of Earth and Space Sciences \& cross-campus Astrobiology Program, University of Washington, Box 351310, Seattle, WA 98195. USA}
\address[add7]{Winton Fellow}

\title{Statistical analysis of Curiosity data shows no evidence for a strong
seasonal cycle of Martian methane}
\date{February 2019}

\begin{abstract}
\setcounter{table}{0}
\renewcommand{\thetable}{5.\arabic{table}}
Using Gaussian Process regression to analyze the Martian surface methane Tunable Laser Spectrometer (TLS) data reported by \citet{Webster2018}, we find that the TLS data, taken as a whole, are not statistically consistent with seasonal variability. The subset of data derived from an enrichment protocol of TLS, if considered in isolation, are equally consistent with either stochastic processes or periodic variability, but the latter does not favour seasonal variation.
\end{abstract}
\end{frontmatter}

\doublespacing

\section{Introduction}
\label{sec:intro}

For the past $\sim$15 years, purported measurements of methane on Mars have elicited excitement \citep{Formisano2004,Krasnopolsky2004,Mumma2009}, puzzlement \citep{Lefevre2009,Lefevre2019} and skepticism \citep{Zahnle2011}. Using methane’s 3.3 $\mu$m absorption band, reports during the years 2004-2019 using remote spectroscopy ranged from non-detections, to purported detections at mean levels $\sim$10-15 ppbv, to claimed enhancements of up to 50 ppbv locally on Mars. The remote techniques have included spectroscopy from ground-based telescopes \citep{Krasnopolsky2004,Krasnopolsky2012, Mumma2009, Villanueva2013} and orbital spectroscopy by the Planetary Fourier Spectrometer (PFS) on the Mars Express orbiter \citep{Formisano2004,Geminale2011, Giuranna2019}. In addition, a weak signal from the 7.7 $\mu$m band in data from the Thermal Emission Spectrometer (TES) on the Mars Global Surveyor orbiter purportedly showed seasonally variable methane at levels of 0-30 ppbv \citep{Fonti2010}. However, the data were revisited and the result was shown to be consistent with a non-detection \citep{Fonti2015}.

Aforementioned reports of remote methane detection and its purported variability have caused controversy \citep{Zahnle2011}. Claims of variability are confronted by the lifetime of methane on Mars of $\sim$300 years, which is based on firmly established chemical kinetic data \citep{Summers2002}. The gas phase photochemistry of methane is well characterized both in the laboratory and in nature because of methane’s importance to the chemistry and climate of the terrestrial atmosphere \citep[e.g.,][]{Burkholder2015, Prather2012}. Thus, claims of ``methane plumes'' that rapidly vanish and methane variability are extraordinary and require extraordinary evidence to be accepted \citep{Lefevre2019, Lefevre2009, Zahnle2011}.

Questions have been raised about the purported detections from TES and PFS because of limited instrumental sensitivity and/or measurement protocols, and the high-resolution remote detection of \citet{Mumma2009} has been attributed to over-nulled $^{13}$CH$_4$ terrestrial emission when purported martian methane lines would have been Doppler blue-shifted onto $\sim$20 times bigger $^{13}$CH$_4$ terrestrial lines \citep{Zahnle2011}. Zahnle et al. claim that when the opposite Doppler shift occurred, putting putative Mars methane lines in a relatively uncontaminated part of the terrestrial spectrum, non-overlapping terrestrial $^{13}$CH$_4$ was modeled correctly (by construction) and nulled out. But see \citet{Villanueva2013} for an alternative view. 

More recently, high-resolution and high signal-to-noise spectral data from the Trace Gas Orbiter (TGO) do not show any methane to a detection limit of 0.05 ppbv \citep{Korablev2019}. With over 100 soundings around the globe, these data are much more numerous than Curiosity Rover data (discussed below). Instruments on TGO – the NOMAD (Nadir and Occultation for MArs Discovery) and ACS (Atmospheric Chemistry Suite)-- were specifically designed to look for trace levels of methane on Mars, and do not suffer from terrestrial contamination. Consequently, it is logical to give the TGO results higher credence than all previous remote spectroscopy measurements.

Before TGO, it was hoped that in situ measurements by the Curiosity Rover TLS (part of the Sample Analysis at Mars (SAM) instrument package), would help resolve debates about martian methane \citep{Webster2011}. Instead TLS measurements continue to generate debate, especially when juxtaposed with the recent TGO non-detection. The TLS instrument has terrestrial methane in its foreoptics chamber, which must be exactly subtracted out in data reduction to derive accurate methane estimates \citep{Webster2013}. Specifically, following pump-outs of the foreoptics chamber on sols 400 and 1000, the methane content in the foreoptics chamber increases, which \citep[][hereafter W18]{Webster2018} explain by ``small amounts of methane released from epoxy and cabling within the chamber'' (W18, Supplemental Material). Background terrestrial methane affects signal-to-noise. Importantly, unlike TLS or ground-based telescopic measurements, the sunlight spectra of TGO instruments during solar occultations by the martian atmosphere are not looking through any terrestrial methane, and so provide sensitive, uncontaminated data \citep{Korablev2018, Korablev2019}.

TLS methane estimates are derived from two measurement protocols: (1) ``direct-ingest'' where intake of martian air is directly measured, and (2) ``enrichment'' where CO$_2$ is removed from the sample of martian air to concentrate the methane and other unreactive gases to provide more sensitive methane measurements \citep{Webster2015, Webster2018}. The enrichment protocol produces an estimated mean background methane level of 0.4 $\pm$ 0.16 ppbv (W18). However, over time, individual methane estimates from the enrichment protocol are scattered in their magnitude. In viewing this scatter, it is claimed by W18 that there is a ``strong seasonal cycle'' in these methane data. Unlike the enrichment protocol results, the direct protocol results provide derived methane estimates that have occasional spikes of $\sim$5-10 ppbv methane (W18).

Establishing whether there is really a seasonal cycle is important because of the conflict between TGO global non-detection to 0.05 ppbv \citep{Korablev2019}, a recent report of a $15.5 \pm 2.5$ ppbv pulse of methane in 2013 from the Mars Express PFS \citep{Giuranna2019}, and results from Curiosity Rover. The purported seasonal cycle of methane on Mars has led to speculation that chemical sinks of methane are occurring, which are unique to Mars and generally near-surface \citep[reviewed by][]{Lefevre2019}, and that this rationalizes the conflict between Curiosity and TGO data. However, it is difficult to see how putative strong, near-surface destruction can be reconciled with the Mars Express or earlier telescopic detections. An attempt to model a seasonal methane cycle uses a destruction timescale for methane that is a free parameter, resulting in methane destruction $\sim$1000 times faster than known chemistry, along with thermodynamic methane absorption significantly different from lab-based measurements \citep{Moores2019}.

Surface chemistry on dust or in soil \citep{Atreya2007} or physical adsorption in soil have generally been invoked to explain methane variations. The former are hypothetical and unique to Mars, while laboratory measurements of the latter \citep{Gough2010}, are insufficient for the required sink \citep{Meslin2011, Hu2016}. Other hypotheses to explain the Curiosity detection of methane include aquifers that release sporadic methane spikes with no seasonality, or seasonal variation due to the presence of life \citep{Hu2016}. Thus,  determining whether methane varies seasonally or not matters for ascertaining the validity of some hypotheses.

The motivation for this study is that the seasonal cycle that has been claimed for the Curiosity methane estimates derived from the enrichment protocol has not been shown to be statistically valid. Here, we examine the time series of the TLS data with statistical methods to see if the data are best explained by stochastic behavior or periodic variability, and, if the latter, whether seasonal cyclicity is favored. W18 stacked all the TLS enriched methane data from different Mars years into one Mars year (their Fig. 1B). However, attempting to demonstrate periodicity by choosing a preferred period {\it a priori} and stacking the data into that period is not statistically robust. The reason is that the data might be better explained by non-periodic stochastic variation or may vary with some other period.

Fortunately, robust statistical techniques to test for periodic signals in sparse data exist and are widely used. For example, such statistical methods are routinely adopted to search for periodic signals in scattered astrophysical data from stars, either to determine the periods of stellar rotation or to infer the presence of orbiting planets. Here, we use Gaussian process (GP) regression, which is a statistical method that has been used in many data-driven fields, e.g., biology, chemistry, physics, finance and data science, amongst others \citep[e.g.,][]{Rasmussen06}.

We apply GP regression to the time series of Curiosity Rover Mars methane estimates to assess whether they show a statistically robust seasonal cycle. For our analysis, we use the values and error bars reported by W18: specifically, globally-inferred values from the enriched protocol that they used to claim seasonality (in their Table 1) and globally-inferred values from the direct protocol (in their Table S2).
	
\section{Methods: Statistical package}
\label{sec:stats}

We wish to evaluate whether the variations in the surface methane level from Curiosity are statistically consistent with periodic or stochastic behavior. The basic dataset consists of a time series of methane abundance estimates, which are limited in number, e.g., only 10 data points for the enriched dataset. An objective, conservative approach begins with no {\it a priori} reason to assume that the methane estimates vary periodically or non-periodically, or stochastically or non-stochastically, i.e., we make no assumptions about which periods or timescales the data are allowed to vary on. This approach allows us to interrogate the data in an unbiased, statistically correct way where the analysis and results are driven by the dataset itself.

Gaussian processes (GPs) offer a flexible framework for performing Bayesian inference on functions. A GP is a non-parametric form of model that defines a distribution over functions. GPs can be thought of as a way to model a dataset by parameterising the covariance between pairs of data points, as opposed to explicitly defining a functional form of model to fit the data. 

As GPs define a distribution over functions, when we vary the parameters of a GP (called hyperparameters), we move through function space rather than the traditional parameter space of parameteric models. GPs have been applied to similar problems before, e.g., assessing the presence of periodicity in the atmospheric CO$_{2}$ concentration measurements made at Mauna Loa, Hawaii \citep{Rasmussen06}. In this analysis, an uninformed model approach was used, where no specific periods (e.g. annual) were weighted a priori, and the GP model found evidence of annual (i.e., seasonal) variation in the CO$_{2}$ level (as well as a long term trend of increasing CO$_{2}$ level with time). This shows that uninformed GP models, which we use here, can infer genuine periodicity in a dataset. A detailed description of GP regression is beyond the scope of this Note, so we refer the interested reader to \citet{Roberts12} and \citet{Rasmussen06} for a general and thorough introduction, respectively, and to \citet{Gillen19} for application to time series of variable stars.

For this work, we use the {\tt Celerite} GP package \citep{Foreman-Mackey17,celerite2} through the {\tt exoplanet} toolkit \citep{exoplanet}\footnote{https://github.com/dfm/exoplanet}. Given the sparsely sampled TLS data and possibility of periodic variation in the methane level on Mars, we opted to use a GP kernel that is akin to a stochastically-driven damped simple harmonic oscillator (SHO). Depending on the level of damping, this kernel can vary smoothly and periodically, or stochastically (with rougher aperiodic variations). It is therefore a powerful tool to assess both the presence and significance of variability within a dataset. Essentially, the SHO GP will explore functions that are consistent with the data: if the data show periodic variability, and are sufficiently constraining, then the GP will favor periodic models. However, if the data show variability, but lack the required level of constraint to assert that it is periodic, then the GP will explore both periodic and aperiodic variability models. Based on the GP posterior distribution, it is possible to assess the level of evidence for or against periodic variability within a dataset, as well as providing an estimate and uncertainty for any such period(s).

To perform posterior inference, we implemented gradient-based Markov-chain Monte Carlo (MCMC) with \textsl{PyMC3} \citep{pymc3} using No U-Turn Sampling \citep[NUTS;][]{Hoffman14}. We ran 5 independent chains of 100,000 steps, which typically yielded around ten thousand effective samples for model evaluation.

\section{Results}
\label{sec:results}

We assess the claim of strong seasonal variation in Mars' background methane by applying the GP regression described above to methane estimates from Curiosity TLS data (W18). We perform three sets of models on the TLS data: 1. considering all data (i.e. direct and enriched); 2. considering the direct data alone; and 3. selecting only the enriched data, which W18 focused on to claim strong seasonal variation. It is important to note that we model the data in time to search for the presence of periodicity in the data, rather than folding the data on an already-determined period and then searching for trends in this phase space. The former allows for unbiased inference while the latter does not.

\begin{figure}
\centering
\begin{subfigure}[b]{0.49\textwidth}
\caption{All data\label{fig:alldata-both}}
\includegraphics[width=\textwidth]{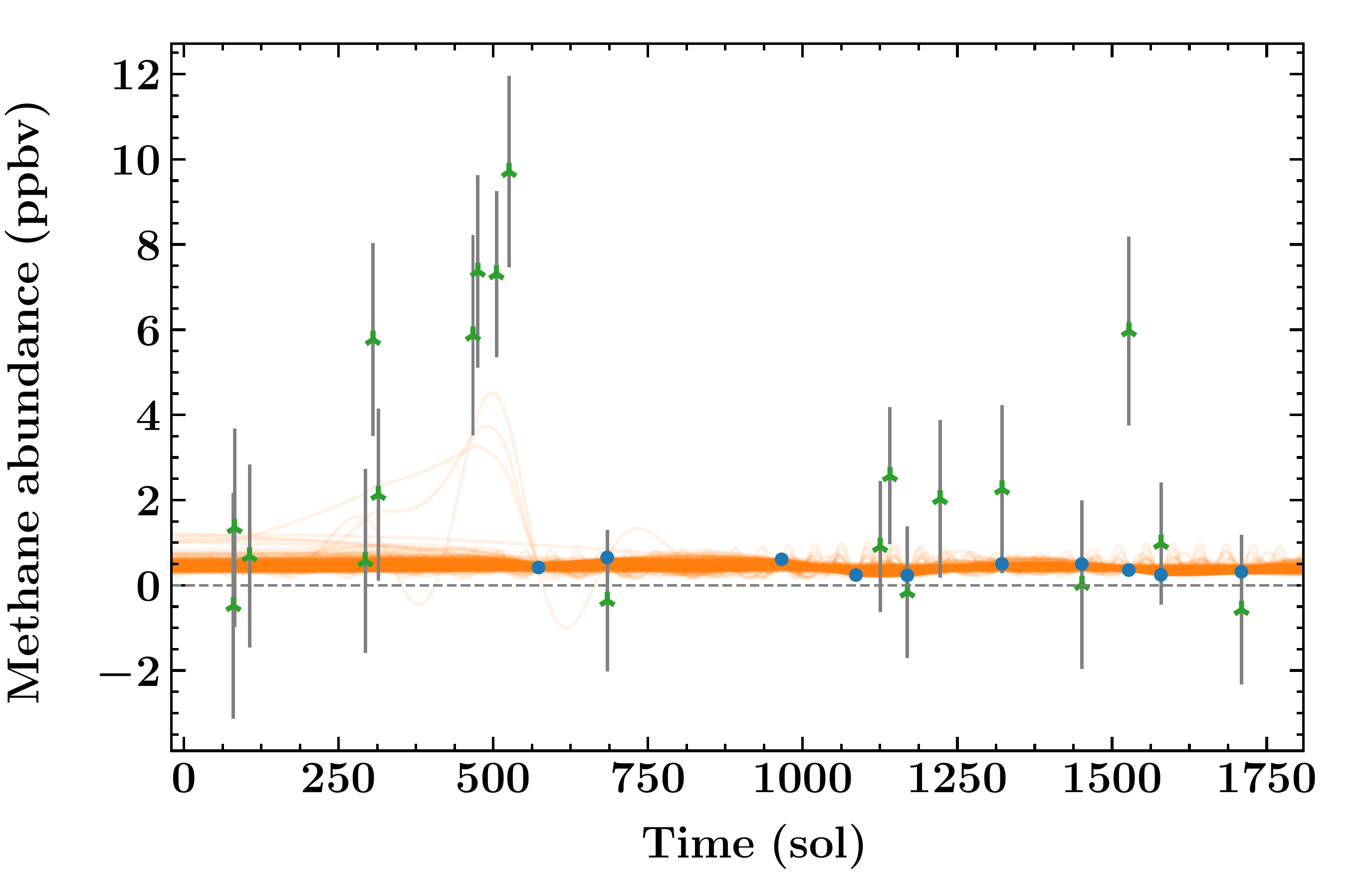}
\end{subfigure}
\begin{subfigure}[b]{0.49\textwidth}
\caption{Enrichment data\label{fig:enrichdata-both}}
\includegraphics[width=\textwidth]{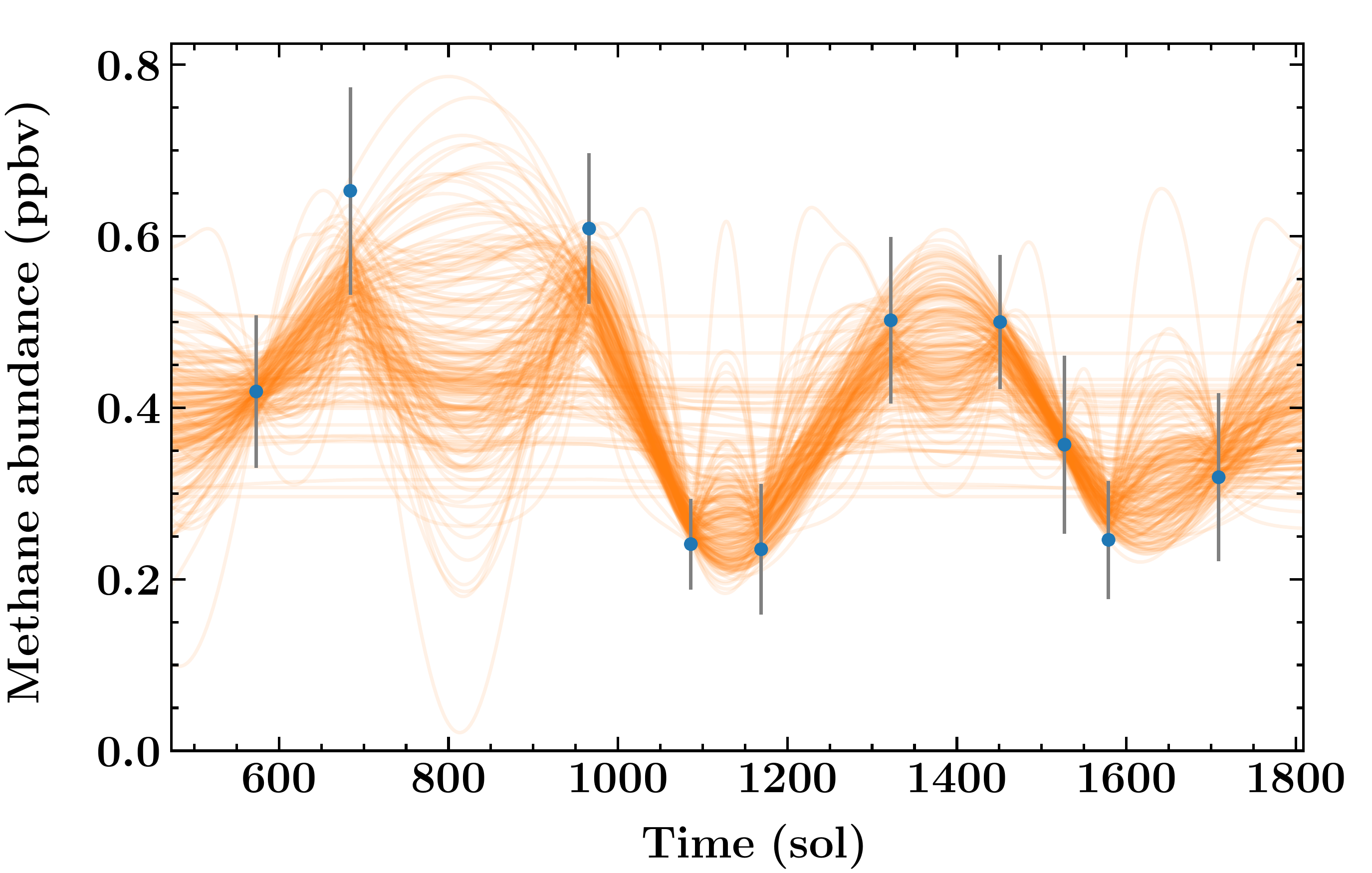}
\end{subfigure}
 
\caption{Methane surface concentration [ppbv] vs. time [sol], using all data (\ref{fig:alldata-both}) and only the enrichment data (\ref{fig:enrichdata-both}), from \citet{Webster2018}, where error bars are 1 standard error. Green triangles represent direct data and blue circles the enriched data. The methane estimates are fit using GP regression, where the orange lines represent 200 individual models drawn from the GP posterior distribution. These give some indication of the range of variability models that are consistent with the sparse TLS data.
\label{fig:data}}
\end{figure}

Fits to both the full and enriched datasets are shown in Figure \ref{fig:data}. Figure \ref{fig:alldata-both} shows 200 realizations of the GP model fit to all data (direct and enriched), which show the kinds of models that are consistent with the data. Taken as a whole, the distribution of models shows no clear periodicity in the background methane, seasonal or otherwise. Figure \ref{fig:enrichdata-both} shows 200 realizations of the GP model fit to only the enriched data, which W18 focused on. We find that many of the individual models do favour relatively smooth variations, but the period of variability is not well constrained due to: 1. the sparseness of the data (10 data points spread over 1136 sols); and 2. the fact that the data do not cover even two full Martian years (1337 sols). This can be seen in the individual draws from the GP posterior distribution, most notably around 800 sol where the data are unable to constrain the GP distribution to a singular form of variation. It is worth noting that we incorporated a white noise jitter term into our GP model. This allows the GP to inflate the observational uncertainties, under penalty, by adding an additional error term in quadrature to the formal error estimates. As stated, any such inflation is penalized by the fit, so the GP will only do this if it is deemed necessary to explain the data, i.e. if the GP finds the formal errors to be underestimated.

\begin{figure}
\centering
\begin{subfigure}[b]{0.49\textwidth}
\caption{All models: periodic and stochastic \label{fig:posterior-all}}
\includegraphics[width=\textwidth]{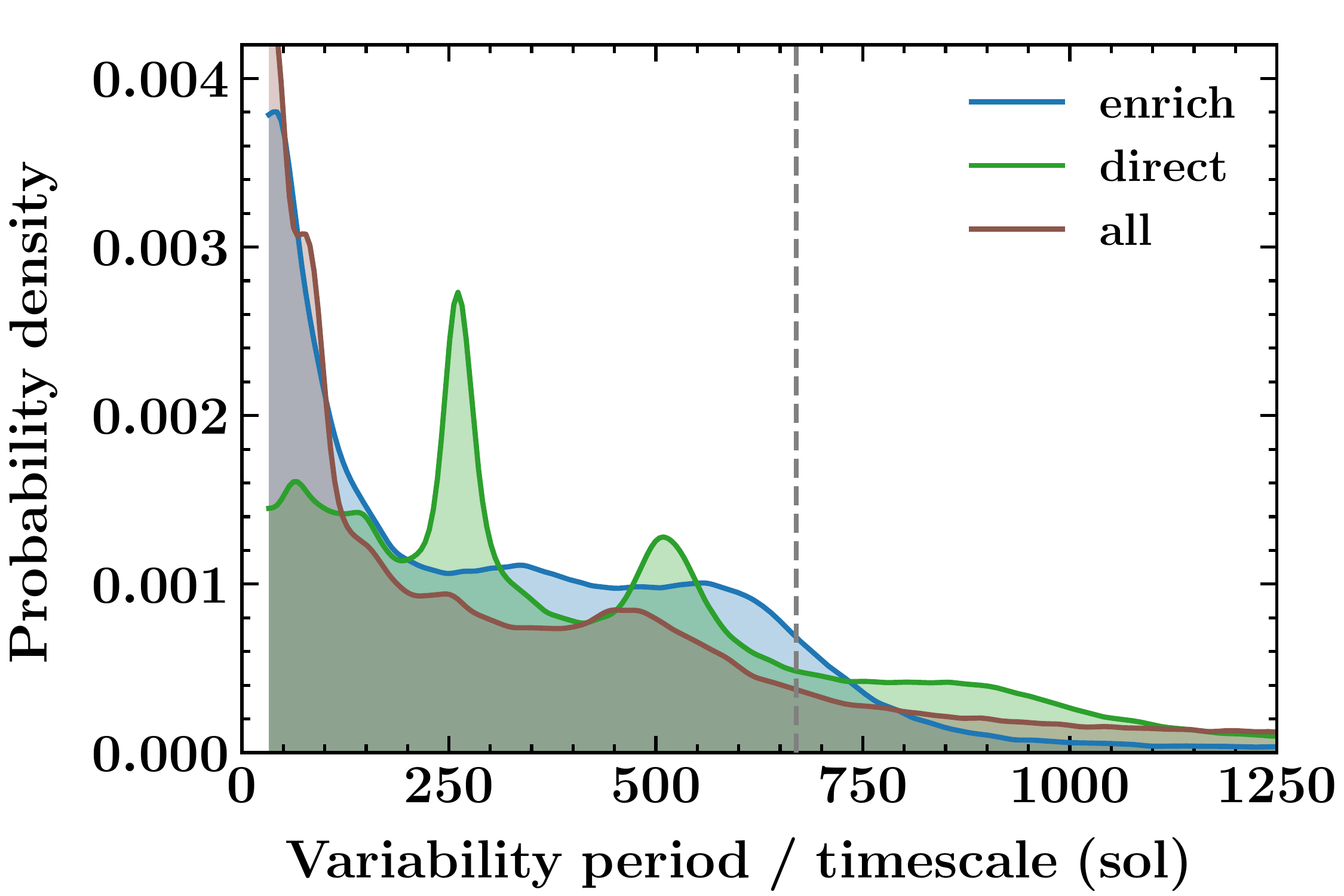}
\end{subfigure}
\begin{subfigure}[b]{0.49\textwidth}
\caption{Only periodic models \label{fig:posterior-periodic}}
\includegraphics[width=\textwidth]{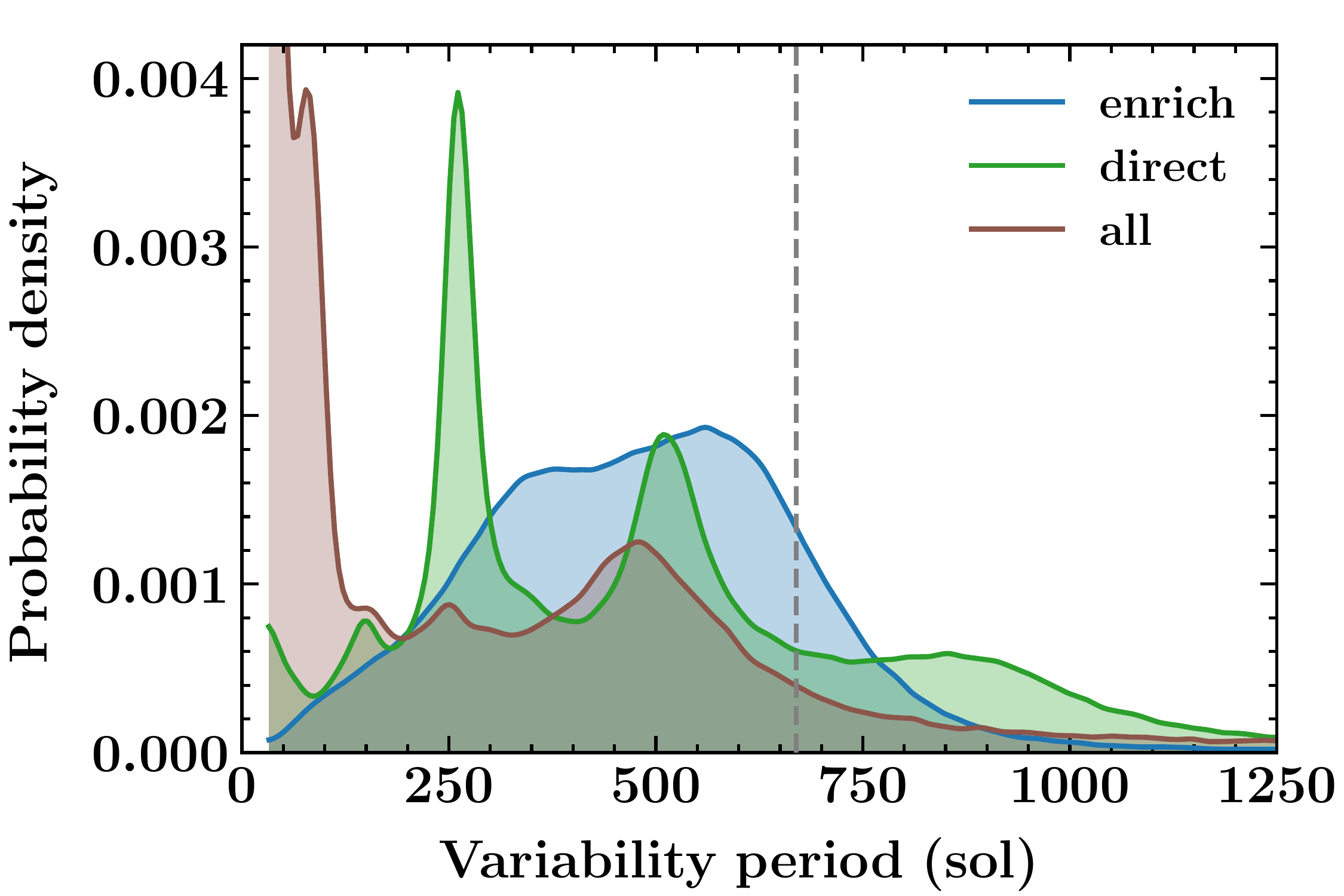}
\end{subfigure}
\caption{Posterior period distributions from the GP model. Higher probability density corresponds to a more strongly favoured period for methane variability. Three posteriors are shown: using all the data (red), only the direct data (green), and only the enriched data (blue). Figure \ref{fig:posterior-all} shows the results when including all GP models and Figure \ref{fig:posterior-periodic} shows the posterior period distribution when selecting only those models which are periodic. The vertical dashed grey line indicates the variability period expected if there were seasonal variation of methane on the Martian surface. \label{fig:posterior}}
\end{figure}

As introduced in \S \ref{sec:stats}, we explore the posterior parameter space of the TLS data using a gradient-based MCMC algorithm. This allows us to efficiently sample a wide range of models, and hence periods, and from this, assess which periods are consistent with the data. The posterior distributions of these consistent periods can then be constructed to help us understand which period(s) are most favoured by the data. Figure \ref{fig:posterior} shows the probability density of models as a function of variability period and/or timescale. We make this distinction because some of the GP models are not strictly periodic and hence we consider their variations to follow a characteristic `timescale' rather than a formal period. Figure \ref{fig:posterior-all} shows the posterior period distribution for all models: this includes both periodic and stochastic models, and hence we explicitly label the horizontal axis as `variability period / timescale'. The distributions show that there are a range of variability periods and timescales consistent with all datasets (enriched, direct and both together), and that variability on the Martian year (668.6 sols) is not preferred over a wide range of other periods or timescales of variation. The top half of Table \ref{tab:posterior_periods} reports the median periods and 1\,$\sigma$  uncertainties for these distributions, which highlight the wide range of allowed periods. We caution the reader not to interpret these values as reliable approximations for the full distributions, which are more complex (e.g., the period distribution for the direct data is multimodal).

In Figure \ref{fig:posterior-periodic} we select only the models that display periodic behaviour, which amounts to $\sim$50$\%$ of the full distributions (see Table \ref{tab:posterior_periods}, bottom half). We note, therefore, that selecting only the periodic models is a strong assumption, which is akin to making a 50:50 guess as to whether the observed variations are periodic in nature or not. Nonetheless, if we assume that periodic variability is in fact present, we again see that variability on the Martian year (i.e. seasonal variation) is not strongly supported by the data. Considering all data together, short periods ($\lesssim$100 sols) are most strongly favoured, but periods up to $\sim$600 sols are plausible. Seasonal variation is not preferred. This is consistent with what W18 suggest for all data: the combined direct and enriched dataset are best explained by stochastic processes. The same can be said of the direct data, which seem to favour periods $\sim$250 or 500 sols, although we note that this is at least in part driven by the sparseness of the data.

When selecting only the enriched data, the posterior distribution in Figure 2b shows that a broad range of variability periods are consistent with the data. While there are a subset of models possessing variability periods that are consistent with the Martian orbital period, this is by no means preferred by the data. Although a seasonal variation, or indeed any variation with a period in the $\sim$100--900 sol range (full range at 10\% maximum), cannot be ruled out, if we consider only periodic models that lie within 5\% (33 sols) of the Martian orbital period, only 9\% of the periodic models are consistent with seasonal variation. Given that only about half of all models are periodic, only 5\% of all models are seasonal with 5\% tolerance on the period; or alternatively, 95\% of models represent either stochastic variation or variation that is on a period other than seasonal. Finally, if periodic variability is present in the enriched data, it is more likely to have a period somewhere between $\sim$300 sols and the Martian orbital period (668.6 sols), than it is to vary on the Martian orbital period itself. 

W18 observe that the sol 965 enriched methane datapoint is higher than their assumed single modal seasonal variation, and speculate that this may be due to a methane spike. In other words, W18 invoke a stochastic process to explain the 965 datapoint, and our model already accounts for stochastic processes. It is inappropriate to run our analysis with the 965 datapoint removed because this point is not a statistical outlier of the enriched dataset: it is less than 0.855 ppbv, which is the third quartile plus 1.5 times the upper interquartile range of the data - the standard statistical rule for a high outlier \citep[][p.38]{Moore2009}. Nonetheless, if the sol 965 datapoint is arbitrarily removed, the result is that half of the models remain stochastic and the fraction of all models that have periods within 5\% ($\sim$33 sols) of a Martian year increases from 5\% to 7\%.

\section{Discussion}
\label{sec:conclusion}

Many models can adequately explain the Curiosity methane data and their parameters cover a large range that must be better constrained by the addition of more surface data before anything definitive about the variability of background methane can be claimed. Given the small amount of data, the values and uncertainties of individual data points become important. In this regard, it is worth noting that reports of TLS CH$_4$ and their uncertainties have seen changes from paper to paper for certain martian sols (e.g., the enrichment factor has changed by a factor of 1.09 between \cite{Webster2015} and W18, while the sol 306 direct in situ value has changed from $-2.21 \pm 0.94$ ppbv (Table 1, Webster et al., 2013) to $5.78 \pm 2.27$ ppbv due to removal of systematic error (Table 1, Webster et al., 2015; Table S2, W18). This suggests that alternative treatments of systematic and random errors in the TLS data could cause further revision of derived CH$_4$ values. This motivated our choice to allow the GP to inflate the observational uncertainties, under penalty, if it deemed necessary for individual models. 

We must wait and see whether future data will reveal seasonal variation in background surface methane, or whether future data, such as future TGO observations in the nadir mode, will better constrain the concentration of surface methane on Mars \citep{Korablev2019}. Here, we simply assess whether the TLS data reported in W18 are evidence favoring the claim of strong seasonal variability over stochastic processes or variability with a different period. We find that they are not.

The sparse Curiosity enrichment data appear by eye to exhibit seasonal variability when the data are phase-folded within a one year period, which effectively weights the data in favor of an annual period, but in-depth statistical analysis is necessary for determining with any confidence whether the data themselves actually favor annual periodicity. Results of statistical analyses show no evidence favoring strong seasonal variation of methane over stochastic variability or variation over a wide range of other periods. The enriched data are roughly equally consistent with either stochastic processes or periodic variability where the latter includes an annual cycle but that cycle is not favored by the data over the many other periods.

More time series data would be desirable because the current enriched dataset comprises 10 measurements spread over 1136 sol, i.e. 1.7 Martian years. To assess whether a periodic signal is present in a dataset with a given period, a dataet covering a timespan of three or more times that specific period is often considered minimal. Furthermore, to reject other potential shorter periods or stochastic variation, a sufficiently high temporal density of measurements is desirable. The exact timespan required and its density of data depend on the complexity of the underlying processes, e.g., whether there is a single well-defined process (e.g., seasonal variation) or whether there are other stochastic factors, and the relative strengths of the various factors.

In conclusion, the statistical analysis in this paper finds that the hypothesis of “strong seasonal variability” in Mars’ surface methane is unsupported by the the Curiosity TLS data. This is because the data are too sparse over too limited a timespan to favor a seasonally cyclic explanation of the data over alternative hypotheses of stochastic variation or variation with other periods.

\section*{Acknowledgments}

We thank Suzanne Aigrain and Kevin Zahnle for helpful comments, and two anonymous reviewers. This paper was initiated while DCC was a Leverhulme Trust Visiting Professor at the University of Cambridge. PBR and DCC thank the Simons Foundation for funding (SCOL awards 599634 to PBR and 511570 to DCC). EG gratefully acknowledges support from the David and Claudia Harding Foundation in the form of a Winton Exoplanet Fellowship.

\pagebreak

\begin{table*}
  \centering
  \caption{Posterior GP variability periods from fitting different combinations of the data: enriched, direct, and both together. Results for the full posterior distribution are shown (top) along with the results from selecting only models that show periodic behaviour (bottom). The \% of the full posterior distribution is indicated for all models highlighting the fraction that show periodic variability. For comparison, note that a Martian year is 668.6 sols (equivalent to 686.98 Earth solar days). We highlight that the posterior period distributions (see Figure \ref{fig:posterior}) are typically complex/multimodal, and hence the median period and 1\,$\sigma$ uncertainty cannot accurately capture the distribution shape. We primarily quote values here to highlight the magnitude of the allowed period ranges.
  } 
  \label{tab:posterior_periods}
  \begin{tabular}{lcc}
    \hline
    \hline
    \noalign{\smallskip}
    Dataset  &  Posterior GP period (Sol)  &  \% of posterior \\
      &  (median $\pm$ 1\,$\sigma$ uncertainty)  & distribution \\
    \noalign{\smallskip}
    \hline
    \noalign{\smallskip}
    
    \multicolumn{3}{c}{......................... Full posterior distributions .........................}\\
    \noalign{\smallskip}
    
    enriched \& direct  &  $363\,^{+1117}_{-290}$  &  100  \\
    \noalign{\smallskip}
    direct              &  $386\,^{+522}_{-249}$   &  100  \\
    \noalign{\smallskip}
    enriched            &  $321\,^{+349}_{-246}$   &  100  \\
    \noalign{\smallskip}
    \noalign{\smallskip}
    \multicolumn{3}{c}{...................... Selecting only periodic models ......................}\\
    \noalign{\smallskip}

    enriched \& direct  &  $341\,^{+588}_{-281}$  &  44  \\
    \noalign{\smallskip}
    direct              &  $491\,^{+414}_{-241}$  &  55  \\
    \noalign{\smallskip}
    enriched            &  $498\,^{+202}_{-205}$  &  47  \\

    \noalign{\smallskip}
    \hline
  \end{tabular}
\end{table*}

\end{document}